\documentclass[%
 aip,
 amsmath,amssymb,
 reprint,%
]{revtex4-1}

\usepackage{graphicx}% Include figure files
\usepackage{dcolumn}% Align table columns on decimal point
\usepackage{bm}% bold math
\usepackage{color}
\usepackage[utf8]{inputenc}
\usepackage[T1]{fontenc}
\usepackage{mathptmx}
\usepackage{etoolbox}

\makeatletter
\def\@email#1#2{%
 \endgroup
 \patchcmd{\titleblock@produce}
  {\frontmatter@RRAPformat}
  {\frontmatter@RRAPformat{\produce@RRAP{*#1\href{mailto:#2}{#2}}}\frontmatter@RRAPformat}
  {}{}
}%
\makeatother
\begin{document}

\preprint{AIP/123-QED}

\title{Tunable topological edge modes in Su-Schrieffer-Heeger arrays}
% Force line breaks with \\
\author{G.~J. Chaplain}
\affiliation{ 
Centre for Metamaterial Research and Innovation, Department of Physics and Astronomy, University of Exeter, Exeter EX4 4QL, United Kingdom%\\This line break forced with \textbackslash\textbackslash
}%
%\email{g.j.chaplain@exeter.ac.uk}
\author{A.~S. Gliozzi}%
 % \email{Second.Author@institution.edu.}
\affiliation{ 
Department of Applied Science and Technology, Politecnico di Torino, 10129 Torino, Italy%\\This line break forced with \textbackslash\textbackslash
}%

\author{B. Davies}
 % \homepage{http://www.Second.institution.edu/~Charlie.Author.}
\affiliation{%
Department of Mathematics, Imperial College London, London SW7 2AZ, UK%\\This line break forced% with \\
}%

\author{D. Urban}
 % \homepage{http://www.Second.institution.edu/~Charlie.Author.}
\affiliation{%
Department of Electronic Systems, Norwegian University of Science and Technology, 7491 Trondheim, Norway%\\This line break forced% with \\
}%

\author{E. Descrovi}
 % \homepage{http://www.Second.institution.edu/~Charlie.Author.}
\affiliation{
Department of Applied Science and Technology, Politecnico di Torino, 10129 Torino, Italy%\\This line break forced% with \\
}%

\author{F. Bosia$^{*}$}
\email[Authors to whom correspondence should be addressed: F. Bosia, federico.bosia@polito.it; R. V. Craster, r.craster@imperial.ac.uk]{ }
 % \homepage{http://www.Second.institution.edu/~Charlie.Author.}
\affiliation{
Department of Applied Science and Technology, Politecnico di Torino, 10129 Torino, Italy%\\This line break forced% with \\
}%

\author{R.~V. Craster$^{*}$}
 % \homepage{http://www.Second.institution.edu/~Charlie.Author.}
\affiliation{%
Department of Mathematics, Imperial College London, London SW7 2AZ, UK%\\This line break forced% with \\
}%

% \date{\today}% It is always \today, today,
%              %  but any date may be explicitly specified

\begin{abstract}
A potential weakness of topological waveguides is that they act on a fixed narrow band of frequencies. However, by 3D printing samples from a photo-responsive polymer, we can obtain a device whose operating frequency can be fine-tuned dynamically using laser excitation. This greatly enhances existing static tunability strategies, typically based on modifying the geometry. We use a version of the classical Su-Schrieffer-Heeger (SSH) model to demonstrate our approach.
\end{abstract}

\maketitle

Topological edge modes perform robust wave localisation at specific frequencies. This makes them an appealing starting point for designing wave control devices. A topological edge mode is formed when symmetries are broken in a periodic system that has a topologically non-trivial band gap, creating an interface along which waves of specific frequencies travel. Due to the underlying band gap material, the waves are unable to propagate away from the interface into the bulk, giving a strong waveguide. The original ``topological insulators'' were developed in quantum mechanics \cite{moore2010birth, hasan2010colloquium,ozawa2019topological}, but the principle has since spread across classical wave physics, including electromagnetism, acoustics, elasticity and mechanics \cite{yang2015topological, rechtsman2013photonic, khanikaev2013photonic, krushynska2023emerging}.

An important feature of topological waveguides is that they are frequency specific. This is either a strength or a weakness (depending on perspective) but it means that it is important that the operating frequency can be tuned to suit the application. There are a variety of tunability strategies in the literature, which often rely on geometric manipulation, achieved for example by manufacturing new samples \cite{chen2016tunable, qu2020minimizing} or physically moving meta-atoms \cite{putley2022tunable}. Tunability can also be achieved by modulating active elements \cite{boardman2011active}. Building on these ideas, we have developed a simple approach for tuning the eigenfrequency of the topological edge mode dynamically and remotely by using a laser to locally alter the material mechanical parameters.
\begin{figure}
    \centering
    \includegraphics[width = 0.35\textwidth]{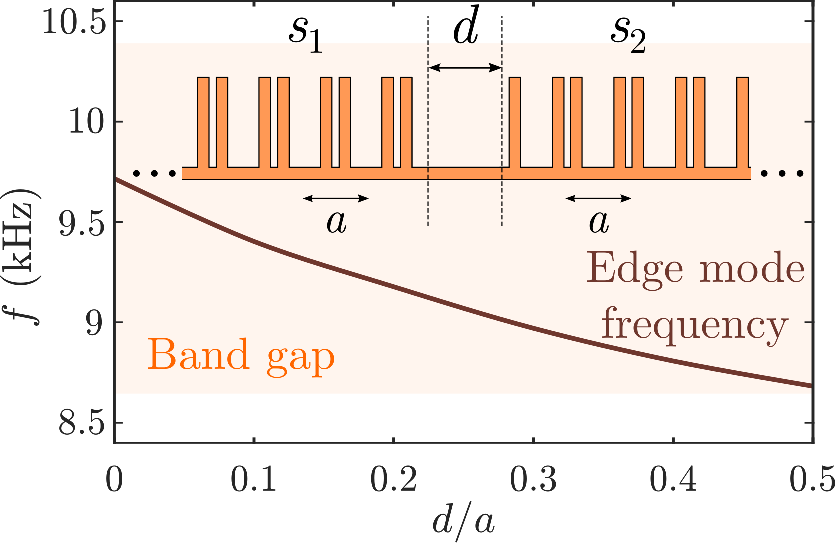}
    \caption{Tuneable edge mode by dislocation: Frequency domain FEM simulations showing change in edge mode frequency with dislocation length $d$ between two SSH chains $s_1$, $s_2$, with unit cell width $a$. We supplement this with dynamic tunability through the use of a photo-responsive material.}
    \label{fig:dislocation}
\end{figure}

We will study a one-dimensional topological system, based on the well-known Su-Schrieffer-Heeger (SSH) model \cite{su1979solitons}, realized with periodically spaced rods on an elastic beam (Fig~\ref{fig:dislocation}). The SSH model is often regarded as the simplest configuration supporting a topologically non-trivial band gap. One-dimensional topological models are not only convenient toy models but also have important applications, for example in energy harvesting \cite{chaplain2020topological} and lasing \cite{parto2018edge}. The basic principle of the SSH model is arranging elements in pairs with alternating `strong' and `weak' coupling strengths (in many classical wave settings, this is realised by alternating `short' and `long' separation distances). In this one-dimensional case, the relevant topological index is the Zak phase \cite{zak1989berry} and edge modes can be created by making a perturbation that creates an interface with different associated Zak phases on either side.

Our approach to tunability has two steps. The first step is to select a geometry that yields an edge mode with the desired eigenfrequency. In the context of the SSH model, this can be done straightforwardly by altering the separation length at the interface, as depicted in Fig~\ref{fig:dislocation}. Tunable topological edge modes based on this principle have been studied theoretically in both one-dimensional \cite{craster2022asymptotic} and three-dimensional \cite{ammari2022robust} models. It was shown that varying the separation length causes the eigenfrequency of the topological edge mode to move across the band gap (Fig~\ref{fig:dislocation}). This idea was motivated by related notions of spectral flow that have been studied in elastic \cite{miniaci2021spectral} and quantum \cite{drouot2020defect} systems.
\begin{figure*}
    \centering
    \includegraphics[width = 0.9\textwidth]{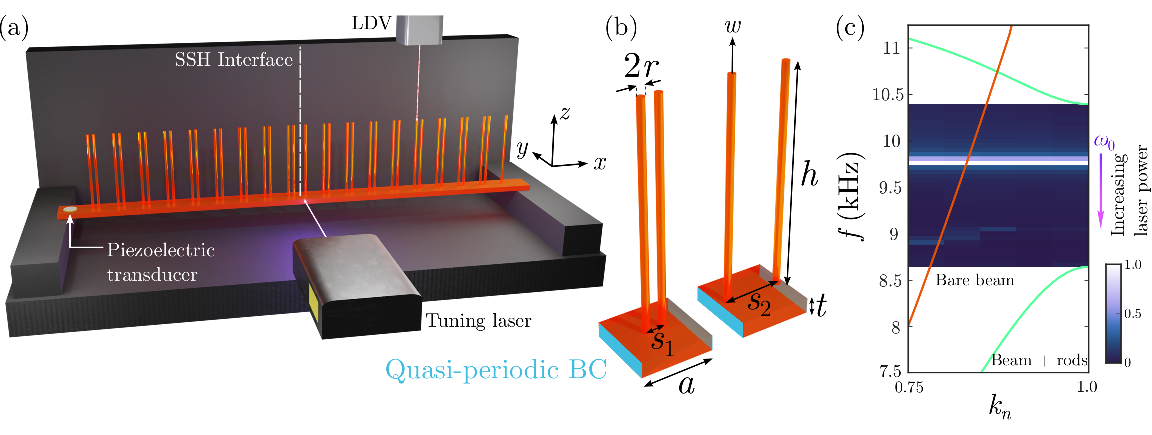}
    \caption{(a) Schematic of experimental setup. (b) Unit cell geometries. (c) Dispersion curves for bare elastic beam (orange) and for the unit cells shown in (b) (green), plotted against normalised wavevector such that $k_n = 1 \equiv k = \pi/a$. Floquet-Bloch quasi-periodic conditions are employed in the FE eigenfrequency model, shown by the blue surfaces in (b). The out-of-plane displacement ($w$) in the $z$-direction (measured in experiments) is also shown. The colourscale in (c) shows normalised Fourier spectra, obtained via time-domain FE simulation (supplementary material), and shows the expected edge mode of frequency $\omega_0$ within bandgap. This frequency $\omega_0$ can be shifted down by increasing the power of the exciting laser.}
    \label{fig:Schematic}
\end{figure*}

The dynamic part of our tunability strategy relies on using photo-responsive polymers for fabrication \cite{gliozzi2020tunable}. Azobenzene-containing dyes are very common molecular photo-switches capable of undergoing a conformational change between two isomeric states upon light absorption \cite{vapaavuori2018supramolecular}. When the two isomer absorption bands have some overlaps, as in this case, photo-isomerization can be cyclic and continuously triggered by single wavelength radiation \cite{poutanen2016structurally}. In such a disordered arrangement of azo-dyes, the photo-thermal effect is then enhanced by photo-isomerization, leading to fast light-induced changes in the polymer mechanical properties \cite{descrovi2018photo}. As detailed below, our samples are 3D printed from a commercial elastic resin that is impregnated with a dispersed azo-dye, producing significant variations of the Young's modulus upon laser illumination. This presents a mechanism for altering the eigenfrequency of topological edge modes. This is valuable, firstly, for correcting some of the effects of any manufacturing imperfections that may undermine our geometric tunability strategy. However, the photo-responsivity is stronger than this as it augments our tunability strategy by adding the possibility for the eigenfrequency to be tuned dynamically.

%\section{SSH interfaces on elastic beams}

We first briefly outline the design of the SSH interface in an elastic wave system consisting of periodically spaced rods on an elastic beam. Such systems have been theoretically shown to support edge modes, with applications in elastic energy harvesting \cite{chaplain2020topological}. A schematic of the full structure, and experimental setup, is shown in Fig.~\ref{fig:Schematic}(a) with the SSH interface marked by a vertical white dashed line. The structure comprises a horizontal elastic beam of thickness $t = 1.5$~mm and width $a = 7$~mm. A periodic array of elastic rods is formed, comprising of a unit cell of width $a$, with two rods of height $h = 20$~mm and radius $r = 0.5$ mm, separated by a distance $s_{i}$ about the cell centre that mediate the `strong' and `weak' couplings (Fig.~\ref{fig:Schematic}(b)). The total length of the support beam is then dictated by the number of unit cells included in the array. We denote the rod separation distance, from the centre of each unit cell, $s_{i}$ where $i = 1,2$ to distinguish between the two sides of the interface. Due to the non-uniqueness of the unit cell, a periodic array with chosen $s_1 = a/3.9$ reproduces the same array as one with a separation $s_{2} = a - s_{1}$, as shown in Fig.~\ref{fig:Schematic}(b). The unit cells are related by a mere translation of $a/2$; they have identical dispersion curves and share a common bandgap. The dispersion relations of each cell are shown in Fig.~\ref{fig:Schematic}(c) (evaluated by an eigenfrequency simulation employing the Finite Element Method (FEM), using COMSOL Multiphysics \cite{comsolSolidMech} - see supplementary material). We form an interface between two arrays (comprising 9 cells) of each configuration, that is, one with central separation $s_1$ and the other with $s_2$. We are guaranteed that this interface supports a topologically protected interface mode by virtue of each configuration sharing a common bandgap in addition to the distinct Zak phase associated with the bands of each geometry \cite{atala2013direct,Xiao2014}. Leveraging the notation of quantum mechanics, the Zak phase for each dispersion curve, $\varphi$, is defined as
 \begin{equation}
     \varphi = i\int\limits_{BZ}\langle u_{\boldsymbol{k}}|\partial_{\boldsymbol{k}}|u_{\boldsymbol{k}}\rangle d\boldsymbol{k}, 
     \label{eq:zak}
 \end{equation}
where $|u_{\boldsymbol{k}}\rangle$ is the wavefunction (eigensolution) of the elastic field at a wavevector $\boldsymbol{k}$ within the first Brillouin Zone (BZ). Several efficient schemes exist to calculate such invariants \cite{Soluyanov2011}. We numerically confirm the existence of this edge mode via a time domain simulation of the experimental setup, shown by the Fourier spectra in Fig.~\ref{fig:Schematic}(c) where the edge mode is labelled $\omega_0$. The introduction of a local defect by altering the material parameter at the interface can fine-tune the frequency of the edge mode. This is achieved by decreasing the Young's modulus, locally, at the interface position on the supporting beam by increasing the control (blue) laser power.

\begin{figure*}
    \centering
    \includegraphics[width = 0.9\textwidth]{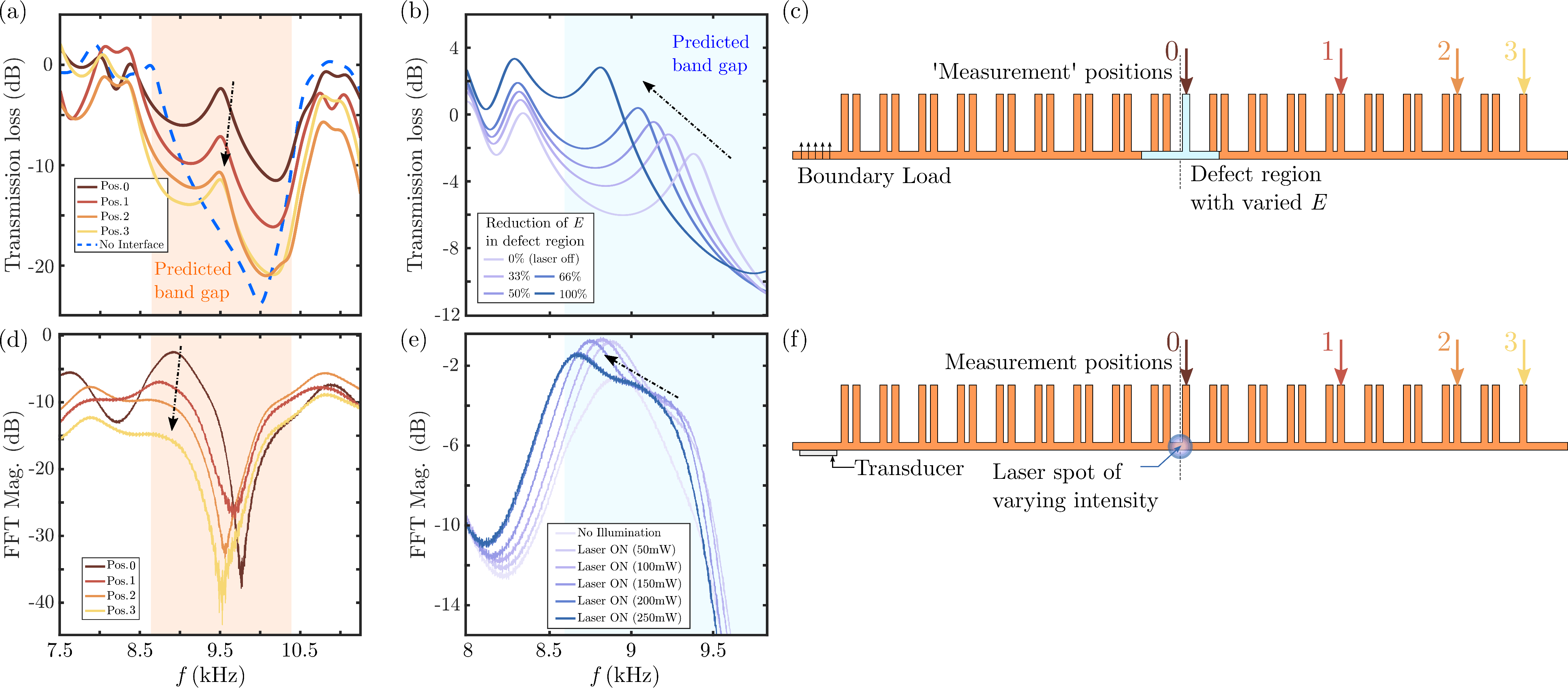}
    \caption{Frequency domain FEM simulations and Experimental results: existence and tunablility of protected edge mode. (a) Numerical confirmation of edge mode shown by the transmission loss as a function of frequency, evaluated at different positions along the sample. Exponential decay with position of the edge mode amplitude is observed away from the interface, as expected. (b) Tuning the frequency of the edge mode by introducing a defect (altered Young's modulus, $E$). (c) Schematic with `measurement' positions (pos. 1-3), defect region shown in blue, and boundary load excitation. (d) Experimental confirmation of edge mode shown by the magnitude of the Fourier spectrum, measured at different positions along the sample. Exponential decay with position of the edge mode amplitude is observed, as expected. (e) Tuning the frequency of the edge mode by introducing a defect with increasing laser power (altering the Young's modulus, $E$). (f) Schematic with measurement positions (pos. 1-3), defect region shown by laser spot blue, and transducer position shown. }
    \label{fig:combined}
\end{figure*}

A more precise estimation of the effect of laser illumination on the edge mode properties can be derived from frequency domain FEM simulations, again performed using COMSOL Multiphysics. A detailed description of the numerical methods is provided in the supplementary material. The sample geometry illustrated in Fig.~\ref{fig:Schematic}(a) is considered, with an overall beam length of 196 mm, corresponding to 28 unit cells of lattice parameter $a$. A harmonic boundary load is applied on the underside of the fourth unit cell from the edge, corresponding to the location of the piezoelectric transducer in Fig.~\ref{fig:Schematic}(a) - shown in Fig.~\ref{fig:combined}(c). The corresponding transmission loss (calculated as the ratio between transmitted and input power for out-of-plane vibrations) is then evaluated at various points along the sample, on top of the rods, in the 5 to 13 kHz range. Adopted material properties are Young's modulus $E$ = 1 GPa, Poisson's ratio $\nu$ = 0.4, density $\rho$ = 1100 kg/m$^3$. Damping is accounted for by introducing an isotropic loss factor of $\eta$ = 0.02. The resulting spectra are shown in Fig.~\ref{fig:combined}(a). Despite being in a band gap (shaded in orange), the localized mode is detected (occurring at approximately 9.5 kHz), and visualized as an isolated peak. We are able numerically (and experimentally) to detect the edge mode with this excitation configuration due to the finite length of the sample coupled with the length of the decay of the mode within the band gap; the decay lengths of frequencies within the gap (which can be predicted through High Frequency Homogenisation \cite{craster2010high, craster2022asymptotic}) reach the interface and ignite the edge mode. In contrast, in Fig. 1(c) we excite in the time domain (as in the experiment) at the interface position (supplementary material). In Fig.~\ref{fig:combined}(a) curves of different colours, representing measurements at different spatial locations, indicate an exponential amplitude decay of the mode when moving from the central defect region ("Pos. 0") to the edge of the sample ("Pos. 3"), as illustrated in Fig.~\ref{fig:combined}(c). The effect of laser illumination on the transmission loss spectrum is shown in Fig.~\ref{fig:combined}(b): a Young's modulus reduction of up to 0.3 GPa ("100$\%$") occurs in the central region of the photo-responsive sample for maximum laser power, and this leads to a frequency downshift of the localized mode of approximately 600 Hz, demonstrating that the tunablility of the edge mode can, theoretically, occur over a significant fraction of the width of the band gap (approx. 35\%).

% \begin{figure}
%     \centering
%     \includegraphics[width = 0.45\textwidth]{NumericalResults2.eps}
%     \caption{Frequency domain FEM simulations: existence and tunablility of edge mode. (a) Confirmation of edge mode shown by the transmission loss as a function of frequency, evaluated at different positions along the sample. Exponential decay with position of the edge mode amplitude is observed, as expected. (b) Tuning the frequency of the edge mode by introducing a defect (altered Young's modulus, $E$). (c) Schematic with 'measurement' positions (pos. 1-3), defect region shown in blue, and boundary load excitation. \textcolor{red}{update schematic with Federico positions -- do we want a zoom in of numerical vs experimental peaks?}}
%     \label{fig:Num}
% \end{figure}

%\section{Experimental Results}

Experimental samples were fabricated using a commercial 3D digital light processing printer (Solflex SF650, W2P Engineering GmbH) with a Solflex Tech polymeric resin (material properties given above). Polymerization was induced with a 385~nm wavelength UV-LED with a power density of 8$\pm$0.5~mW/cm$^2$. The nominal lateral resolution is 50~$\mu$m and printing was performed in layers of 50~$\mu$m, each exposed for 1.5~s, leading to a dose of 12~mJ/cm$^2$ per layer. After printing, the sample was washed with Isopropyl Alcohol (IPA), sonicated in IPA, and dried thoroughly. It was then soaked for 10 minutes in an acetone bath containing 0.25 mg/mL photo-responsive azo-dye Disperse Red 1 methacrylate (DR1m), swelling the pre-cured structure, and thus permitting homogeneous dye dispersion. Finally the swollen structure was dried for 1 hour until the pristine shape was recovered before being UV-post-cured for 1 hour under all-around illumination in the wavelength range 320–450~nm.

The experimental setup is shown in Fig.~\ref{fig:Schematic}(a): a 20~mm diameter piezoelectric transducer was glued to the sample at one end and a frequency sweep between 5~kHz and 13~kHz (with a duration of 1~s) was injected into the sample using an Agilent 33500 function generator, amplified by a linear amplifier (FLC Electronics, A400DI). The transducer was glued to the underside of the sample, but is shown on the top side of Fig.~\ref{fig:Schematic}(a) for clarity. The out-of-plane ($z$-direction in Fig.~\ref{fig:Schematic}) vibration velocity amplitude was measured at various points along the sample, on top of the rods, using a Polytec OFV-505 laser vibrometer. Using an additional 405~nm-wavelength laser beam impinging laterally on the central region of the sample (see Fig.~\ref{fig:Schematic}(a)), the material's Young's modulus can be reduced locally by light stimulation. The laser has a maximum power output of 250~mW and can be focused on an approximately 30~mm$^2$ area. 

The Fast Fourier Transform (FFT) of the acquired signals is plotted in Fig.~\ref{fig:combined}(d), clearly showing the band gap region and an isolated peak corresponding to the localized mode near 9 kHz. As the signal detection point moves from the central position 0 to the lateral position 3 (Fig.~\ref{fig:combined}(f)), the FFT peak amplitude drops exponentially, as predicted theoretically and in simulations, while other spectral features remain relatively unchanged. The experimental peak widths appear larger than those in simulations, due to an oversimplified viscoelastic model which neglects significant damping. Illumination of the central defect region with the laser, for increasing values of output power, is shown in Fig.~\ref{fig:combined}(e), demonstrating that the peak relative to the localized mode shifts to lower frequencies, approaching the band edge while reaching a maximum variation of approximately 300 Hz for 250 mW, compared to the original frequency. The variation is qualitatively similar to the predictions by theory and simulation, but less drastic due to the fact that the adopted laser power is not sufficient to generate a full photo-isomerization of the azo-units in the polymeric material. The peak shift is fully reversible when removing illumination or decreasing illumination power in any number of cycles. Similar results were obtained on a second specimen with different geometric parameters (see Supplementary Material).    

% \begin{figure}
%     \centering
%     \includegraphics[width = \textwidth]{ExperimetalResults2.eps}
%     \caption{Experimental results: existence and tunablility of edge mode. (a) Confirmation of edge mode shown by the magnitude of the Fourier spectrum, measured at different positions along the sample. Exponential decay with position of the edge mode amplitude is observed, as expected. (b) Tuning the frequency of the edge mode by introducing a defect with increasing laser power (altering the Young's modulus, $E$). (c) Schematic with measurement positions (pos. 1-3), defect region shown by laser spot blue, and transducer position shown. \textcolor{red}{update schematic with Federico positions}}
%     \label{fig:Exp}
% \end{figure}

%\section{Summary and Outlook}

In summary, we have demonstrated experimentally the possibility of dynamically fine tuning the frequency of topological edge modes in SSH arrays consisting of appropriately spaced rods on a beam, realized in a 3D-printed photo-responsive polymer. The frequency shift due to local illumination of the sample is linearly dependent on laser power and can reach the band gap edge, and is fully reversible when illumination is removed. Measurements have been performed on two different sample geometries and provide similar results, demonstrating flexibility in design possibilities.        

This work shows the potential to use photo-responsive materials to create topological edge modes with dynamic tunability. This builds on existing strategies for tunability through geometric manipulation. This principle could be developed to multi-dimensional topological systems, to add dynamic tunability to topological waveguides and energy splitters. By providing a means to dynamically alter the operating frequency, this would greatly increase the the versatility of these devices.

%\section*{Supplementary Material}
%See Supplemental Material at [URL
%will be inserted by publisher] for details on frequency and time domain finite element models, and %an additional sample geometry.

\section*{Supplementary Material}
\noindent Details of the FE modelling, along with an additional sample geometry, are presented in the supplementary material.

\begin{acknowledgments}
A.S.G., F.B. and R.V.C. are supported by the European Commission H2020 FET Open “Boheme” grant no. 863179. B.D. is supported by the Engineering and Physical Sciences Research Council through a Research Fellowship with grant number EP/X027422/1. G.J.C. gratefully acknowledges financial support from the Royal Commission for the Exhibition of 1851 in the form of a Research Fellowship.
\end{acknowledgments}

\section*{Data Availability Statement}
The data that support the findings of this study are available from the corresponding author upon reasonable request.

\end{document}